\documentclass{emulateapj}
\usepackage{rotating}

\newcommand{\vecg}{\mbox{\boldmath $g$} {}}

\shorttitle{Slowly-rotating galaxies}
\begin{document}
\title{Slowly rotating gas-rich galaxies in modified Newtonian dynamics (MOND)}

\author{F.~J.~S\'anchez-Salcedo\altaffilmark{1},
          A.~M.~Hidalgo-G\'amez\altaffilmark{2} and 
          Eric E.~Mart\'{\i}nez-Garc\'{\i}a\altaffilmark{1}}
\altaffiltext{1}{Instituto de Astronom\'ia, Universidad Nacional Aut\'onoma
de M\'exico, Ciudad Universitaria, 04510 Mexico City, Mexico;
jsanchez@astro.unam.mx}
\altaffiltext{2}{Departamento de F\'{\i}sica, Escuela Superior de F\'{\i}sica
y Matem\'aticas, IPN, U.P. Adolfo L\'opez Mateos,
C.P. 07738, Mexico City, Mexico}

\begin{abstract}
We have carried out a search for gas-rich dwarf galaxies that have
lower rotation velocities in their outskirts
than MOdified Newtonian Dynamics (MOND) predicts, so that the amplitude
of their rotation curves cannot be fitted by arbitrarily increasing
the mass-to-light ratio of the stellar component or by assuming
additional undetected matter.
With presently available data, the gas-rich galaxies 
UGC 4173, Holmberg II, ESO 245-G05, 
NGC 4861 and ESO 364-G029 deviate most from MOND predictions
and, thereby, provide a sample of promising targets 
in testing the MOND framework. 
In the case of Holmberg II and NGC 4861,
we find that their rotation curves 
are probably inconsistent with MOND, unless
their inclinations and distances differ significantly from 
the nominal ones.
The galaxy ESO 364-G029 is a promising target because its
baryonic mass and rotation curve are similar to Holmberg II but 
presents a higher inclination. Deeper photometric and H\,{\sc i} 
observations of ESO 364-G029, together with further decreasing systematic
uncertainties, may provide a strong test to MOND.
\end{abstract}


\section{Introduction}
Milgrom (1983) proposed that a modification of Newton gravitational
law at accelerations below a threshold of $\approx 10^{-8}$ cm
s$^{-2}$, could explain the dynamics of galaxies without invoking
any dark matter.
In the so-called MOdified Newtonian Dynamics (MOND), the rotation
velocity of an isolated galaxy is determined by its visible (baryonic)
mass. MOND has proven successful in reproducing
the shape and amplitude of a significant fraction of
spiral galaxies (without any dark matter) with only the mass-to-light
ratio of the stellar disk as adjustable parameter (see Sanders \&
McGaugh 2002 for a review; Milgrom \& Sanders 2007; Sanders \& Noordermeer 
2007; Swaters et al.~2010).

In this paper we will consider gas-rich dwarf galaxies to test MOND.
This type of galaxies provide a good test for MOND because (1) in most
of these galaxies the internal acceleration is below the threshold
and (2) their mass is dominated by gas and hence the predicted rotation
curve is not sensitive to the assumed stellar mass-to-light ratio 
$\Upsilon_{\star}$.
In the MOND framework, the asymptotic velocity, that is, the rotation
velocity in the outermost regions where rotation curves tend to be flat,  
is $(GMa_{0})^{1/4}$, where $M$ is the total baryonic mass of the
system. 
For a sample of $47$ gas rich galaxies selected by a strict
criteria, McGaugh (2012) finds a tight empirical relation between 
detected baryonic mass and rotation velocity in the outermost measured 
regions, the baryonic Tully-Fisher
(BTF) relation, as that predicted by MOND.  
The low scatter of the data points relative to their error bars
led McGaugh (2012) to argue that all deviations from the BTF line
can be explained by observational uncertainties alone.
This provides a check of MOND
with zero free parameters [but see also Foreman \& Scott (2011) and 
the reply by McGaugh (2011)]. However, we must note that  
it was already known at that time that the rotation curves of 
$22$ of the galaxies in McGaugh's sample ($47\%$ of the galaxies)
were in agreement with MOND predictions and, hence, they
were not really new.

An intrinsic scatter in the BTF relation would be difficult to accommodate
in MOND because, in this theory, the BTF relation is a direct
consequence of the effective force law. 
On the contrary, the $\Lambda$CDM paradigm predicts a real scatter
albeit small (Desmond 2012). Hence, the BTF relation can in principle
be used to distinguish between both scenarios. 
For now, it is unclear whether
the small scatter in the BTF relation is a proof of MOND or it is
a selection effect.
Following this line of arguments, we have carried out a search for gas 
rich galaxies that separate from the BTF relation.  
If MOND is correct, all these galaxies should have an obvious
problem with the observed rotation curve like uncertain inclinations and
distances, or the presence of non-circular motions.
In particular, we are interested in galaxies that rotate more slowly
than MOND predicts, so that the amplitude of their
rotation curves cannot be fitted by arbitrarily increasing the 
mass-to-light ratio of the stellar component or by assuming additional
undetected matter. 
The selected galaxies might also provide a good target for studying the issue 
of a link between the level of stability of galaxies in MOND and the
star formation (S\'anchez-Salcedo \& Hidalgo-G\'amez 1999).

The paper is organized as follows. In \S \ref{sec:basic}, we give
a brief description of MOND in galaxies. The selection procedure 
of galaxies and the outcome of the search
are presented in \S \ref{sec:search}. Section \ref{sec:testcases}
contains a more exhaustive study focusing
on some test cases. 
Conclusions are given in \S \ref{sec:conclusions}.

\section{MOND: Basic equations}
\label{sec:basic}
The Lagragian MOND field equations lead to a modified version of Poisson's equation
given by
\begin{equation}
{\mbox{\boldmath $\nabla$}}\cdot 
\left[\mu\left(\frac{|{\mbox{\boldmath $\nabla$}}\Phi|}{a_{0}}\right)
{\mbox{\boldmath $\nabla$}}\Phi\right] =4\pi G \rho,
\end{equation}
where $\rho$ is the density distribution, $\Phi$ the gravitational
potential, $a_{0}$ is a universal acceleration of the order of 
$10^{-8}$ cm s$^{-2}$. The interpolating function
$\mu(x)$, with $x=|{\mbox{\boldmath $\nabla$}}\Phi|/a_{0}$, has 
the property that $\mu(x)=x$ for $x\ll 1$ and $\mu(x)=1$ for $x\gg 1$ (Bekenstein \& Milgrom 1984).
Brada \& Milgrom (1995) showed that, to a good approximation, the {\it real} acceleration at
the midplane of an isolated, flattened axisymmetrical
system, $\vecg$, is related with
the Newtonian acceleration, $\vecg_{N}$, by:
\begin{equation}
\mu\left(\frac{|\vecg|}{a_{0}}\right)\vecg=\vecg_{N}.
\label{eq:algMOND}
\end{equation}
The two most popular choices for the interpolating function are
the ``simple'' $\mu$-function, suggested by Famaey \& Binney (2005),
\begin{equation}
\mu(x)=\frac{x}{1+x},
\end{equation}
and the ``standard'' $\mu$-function
\begin{equation}
\mu(x)=\frac{x}{\sqrt{1+x^{2}}},
\end{equation}
proposed by Milgrom (1983). 
Unless otherwise specified, we will take $a_{0}=1.2\times 10^{-8}$ cm s$^{-2}$
and use the simple $\mu$-function (Famaey et al.~2007; Sanders \& 
Noordermeer 2007; Weijmans et al.~2008). 
Since we are interested in dwarf galaxies whose dynamics
lies in the deep MOND regime (that is, $x=g/a_{0}\ll 1$), our results 
are not sensitive to the exact form of the interpolating function.

\section{Gas-rich dwarf galaxies with ``low rotation''}
\label{sec:search}
Define $\Gamma(R)$ as the ratio between the real acceleration
$g$ and the Newtonian acceleration $g_{N}$ 
at galactocentric radius $R$, that is, 
$\Gamma(R)=g/g_{N}=v_{c}^{2}/(v_{c,\ast}^2+v_{c,g}^{2})$, 
where $v_{c}$ is the observed circular and 
$v_{c,\star}$ and $v_{c,g}$ are the Newtonian contributions
of the stars and the gas to the rotation curve, respectively.
At large enough galactocentric
radii, $\Gamma$ is a measure of the ratio between the 
dynamical and the baryonic mass. 
According to Eq.̣~(\ref{eq:algMOND}), if we know $v_{c}(R)$, 
MOND predicts $\Gamma (R)$ as 
$\Gamma_{M}=1/\mu(g/a_{0})$, where $g=v_{c}^2/R$. 
The predicted value $\Gamma_{M}$ can be confronted with the observed
value $\Gamma_{\rm obs}$.
If MOND is correct, $\Gamma_{M}=\Gamma_{\rm obs}$.

In the outer parts of dwarf galaxies,
MOND predicts high $\Gamma$-values for galaxies with low rotation velocities. 
For instance, for a galaxy
with $v_{c}\simeq 40$ km s$^{-1}$ at a radius of $7$ kpc, MOND predicts
$\Gamma_{M}\approx 15$.
In the present study, we will evaluate $\Gamma_{\rm obs}$ in the outer parts
of gas-rich galaxies and select those having the smallest values
of $\Gamma_{\rm obs}$. Doing so, we will able to identify potential
galaxies for which MOND could fail to reproduce the observed amplitude
of their rotation curves. The selected galaxies will deserve a detailed
analysis of all the available rotation curve data. 

To be completely successful, MOND should account for not only the amplitude
of the rotation curves at the outskirt of galactic disks 
but also the detailed shape of the rotation curves. 
However, a test of MOND based on the capability of reproducing the 
fine structure of
the rotation curves is a more delicate issue. Indeed, there exists 
a handful of galaxies that MOND does not provide a good fit
to the shape of the rotation curve (e.g., Lake \& Skillman 1989;
Bottema et al.~2002; S\'anchez-Salcedo \& Lora 2005;
Corbelli \& Salucci 2007; Swaters et al.~2010).
However, this fact does not necessarily rule out MOND
because of uncertainties in distance and inclination
of the galaxies, beam smearing, non-circular motions,
morphological asymmetries, corrections for asymmetric drift, warps and 
uncertainties in the photometric calibration (e.g., Swaters et al.~2010).
Moreover, the simple relation between $g$ and $g_{N}$ given in 
Eq.~(\ref{eq:algMOND})
has only been tested with the underlying assumption of axisymmetry
and may induce some error when estimating the MOND-predicted rotation curves
in non-axisymmetric galaxies.

Another source of uncertainty in the exact shape of the predicted rotation
curve is the specific form adopted for $\mu$, which is important
at intermediate galactic radii, where the transition between
the Newtonian and MOND regimes takes place.
In our selection procedure,  we will evaluate $\Gamma_{\rm obs}$ 
and $\Gamma_{M}$ at the outer parts, where
the predicted $\Gamma_{M}$-value is not sensitive
to the exact form of the interpolating function $\mu$.

\subsection{Selection procedure and computation of $\Gamma_{\rm obs}$
}
In order to check if all galaxies with `slow rotation' present large
values of $\Gamma$,
we have estimated $\Gamma_{\rm obs}$ 
for more than $80$ galaxies with published rotation curves
whose rotation velocities, at the last measured point, 
are between $35$ and $80$ km s$^{-1}$.
Galaxies with rising rotation curves in the outermost measured regions
were also included.
We have discarded galaxies
with circular velocities $<35$ km s$^{-1}$ at the last measured-point to
avoid the inclusion of rotation curves with large uncertainties due to
asymmetric drift corrections (e.g., IC 1613 and UGC 7577). 
On the other hand, we restrict to galaxies with circular velocities 
$<80$ km s$^{-1}$ because they 
are more likely to be gas-rich galaxies (e.g., McGaugh 2012) and
their dynamics lie in the deep MOND regime.
We selected those galaxies with low $\Gamma$-values, 
i.e.~galaxies with $\Gamma_{\rm obs}<5.5$. 

In order to estimate $v_{c,g}$, the gas mass was taken as $1.4$ times 
the H\,{\sc i} content to
correct for the presence of He and metals. The typical uncertainty in the 
H\,{\sc i} mass is less than $5\%$ (except for E364-G029 which is
of $25\%$). Hence, most of the uncertainties in $v_{c,g}$ comes
from the content of molecular gas and other forms of gas. 
In order to estimate $v_{c,\ast}$, the stellar disk contribution to the 
rotation curve, we convert luminosity
to mass using the models of Bell \& de Jong (2001).
The $(B-R)$ color was determined from the magnitudes of the galaxies as
given in NED.  When available, the $25^{th}$
isophote magnitude was preferred. 
In Column 7 of Table \ref{table:parameters} we give a range
of the stellar mass-to-light ratios in the blue band $\Upsilon_{\star}^{B}$.
The range is so ample that
it includes values previously reported by other authors, as well
as 
the bursting model, which is the model that best describes most
of the properties of these galaxies. 
Note that the stellar mass-to-light ratios are very dependent on
the color and on the model.
However, since the gas mass dominates in these galaxies, uncertainties
in the stellar mass-to-light ratio do not have a strong impact on the estimates
of the total baryonic mass or in $\Gamma$
(see columns 8 and 10 in Table \ref{table:parameters}).

\subsection{Search outcome: galaxies with low $\Gamma_{\rm obs}$}
\subsubsection{General comments}
\begin{figure*}
  \includegraphics[width=160mm,height=65mm]{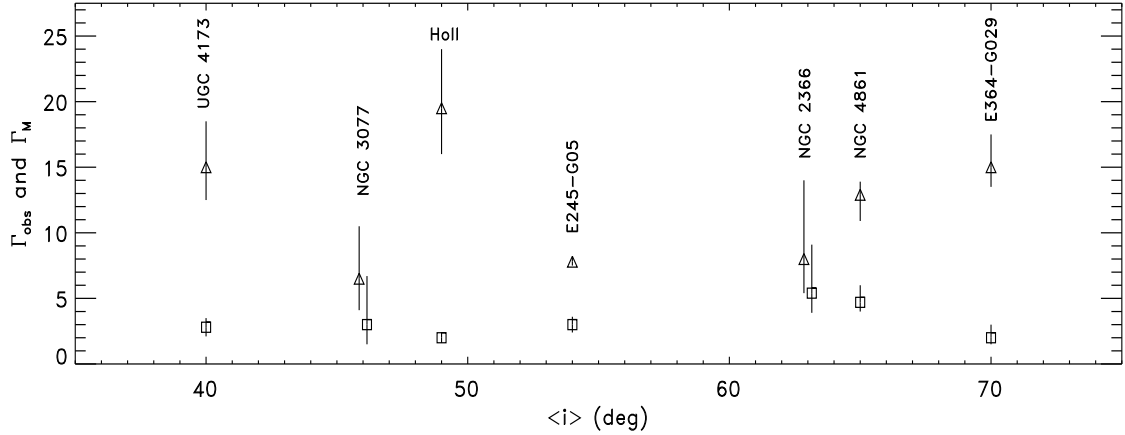}
\vskip 0.4cm
  \caption{
$\Gamma_{\rm obs}$ (squares) and $\Gamma_{M}$ (triangles)
for the selected galaxies.
The galaxies were ordered according to their mean inclination (abscissa).
  }
  \label{fig:gamma_incl}
\end{figure*}

The search turned out only seven gas-rich galaxies with $\Gamma_{\rm obs}<5.5$:
NGC 3077, NGC 2366, UGC 4173, NGC 4861, ESO 245-G05, 
ESO 364-G029 and HoII.
The main properties of these galaxies are compiled 
in Table \ref{table:parameters}; the rotational velocities
at the outer parts, $V_{\rm rot}$, are given in
column 9, the empirical values of $\Gamma$ for 
the seven selected galaxies are given in column 10 (denoted by 
$\Gamma_{\rm obs}$),
whereas the values predicted by MOND, $\Gamma_{M}$, are provided in 
column 11.
The uncertainty in $\Upsilon_{\star}^{B}$ was also treated
as an additional source of error. However, the most important error 
source in the quoted values of $\Gamma$ is due to uncertainties
in the H\,{\sc i} rotation curve of these galaxies.
Figure \ref{fig:gamma_incl} shows the predicted MOND
$\Gamma$-value and the observed one for these galaxies. 

For NGC 3077 and NGC 2366, the values of $\Gamma_{\rm obs}$ and 
$\Gamma_{M}$ are consistent within the $1\sigma$ uncertainty.
The H\,{\sc i} kinematics of NGC 3077 is highly perturbed by the tidal
interaction with M81 and M82 and, therefore, is not a good canditate 
to test MOND. On the other hand, whereas the inclination of NGC 2366 
($i=63^{\circ}$)
is adequate to accurately estimate the rotation curve, the uncertainties 
in the rotation curve are very large. Therefore, these
two galaxies should be discarded until more precise data is available. 
For the remainder five galaxies,
MOND predicts too large a value of $\Gamma$ 
and, hence, it appears to fail for those galaxies with small $\Gamma_{\rm obs}$.

Note that the derived values of $\Gamma_{\rm obs}$ and $\Gamma_{M}$ 
depend on the distance to the galaxy and on the adopted inclination.
The error bars quoted in Table \ref{table:parameters}
do not include uncertainties
in the inclination angle or distance to the galaxy.

Consider first uncertainties in the inclination angle.
Beyond a few disk scale radii, where the mass distribution has essentially
converged, $v_{c,\star}$ and $v_{c,g}$ do not depend on the adopted
inclination. 
The amplitude of the rotation curve at the outer parts $V_{\rm rot}$, 
depends on the inclination $i$ of galaxy and
changes as $V_{\rm rot}\propto (1/\sin i)$.
For modest changes in the inclination so that the outer dynamics
lie in the deep MOND regime, 
it holds that $\Gamma_{M}\propto V_{\rm rot}^{-2}\propto \sin^{2}i$,
whereas $\Gamma_{\rm obs}\propto V_{\rm rot}^2 \propto
(1/\sin^{2}i)$. 

In order to avoid strong effects on the determination
of the rotation curve, Begeman et al.~(1991) required that the galaxies
should have H\,{\sc i} inclinations between $50^{\circ}$ and $80^{\circ}$.
On the other hand, McGaugh (2012) demands consistency between optical
and H\,{\sc i} inclinations.
Whereas ESO 245-G05, NGC 4861 and ESO 364-G029 have comfortable
H\,{\sc i} inclination angles between $50^{\circ}$ and $80^{\circ}$,
none of the selected galaxies comply with McGaugh's
consistency criterion.

The galaxy ESO 215?G009 exemplifies the importance of having galaxies with
high inclinations in order to ameliorate the uncertainties in this
parameter. 
Warren et al.~(2004) report 
a circular velocity of $51\pm 8$ km s$^{-1}$ at $7$ kpc 
for an inclination $i=36^{\circ}\pm 10^{\circ}$. However, 
van Eymeren et al.~(2009a), using
a smaller inclination of $28^{\circ}$, derive a remarkably 
larger amplitude of the
rotation curve of $75$ km s$^{-1}$ at the same angular distance. 
Given the low inclination and the
uncertainties in the distance to this galaxy,
the amplitude of the rotation curve may be well in accordance to 
MOND predictions. Hence, ESO 215?G009 must be excluded.

Uncertainties in the distance $D$ may also play a role.
Consider now how $\Gamma_{\rm obs}$ depends on the adopted distance
to the galaxy.
Assuming that the H\,{\sc i} disk is infinitelly thin, $v_{c,g}^2\propto D$.
The stellar contribution $v_{c,\ast}^{2}$ is also proportional
to $D$ if the stellar mass-to-light ratio is kept fixed.
Therefore, $\Gamma_{\rm obs}\propto (v_{c,g}^2+v_{c,\ast}^2)^{-1}\propto
D^{-1}$.
The variation of $\Gamma_{M}$ to changes in the adopted distance
is also simple to derive. 
Suppose a galaxy that rotates at velocity $V_{\rm rot}$ at a radius $r_{m}$.
In the deep MOND regime, we have that 
$\mu\simeq g/a_{0}=V_{\rm rot}^{2}/(r_{m}a_{0})$ at $r_{m}$.
Since $r_{m}$ scales as $r_{m}\propto D$, we obtain 
$\Gamma_{M}=1/\mu \propto r_{m}\propto D$ (provided that the
distance change is small enough that the outer galaxy is still
in the deep MOND regime).

Can uncertainties inherent to such a sample of
galaxies like uncertainties in the inclinations and distances 
explain the discrepancy between $\Gamma_{\rm obs}$ and $\Gamma_{M}$?
In the following we will try to answer this question.
Before making any further interpretation, it is convenient
to briefly comment on individual galaxies.

\subsubsection{Comments on individual galaxies}

The MOND rotation curve fitting of UGC 4173 was already
studied by Swaters et al.~(2010). Using the standard $\mu$-function,
MOND overestimates the circular velocity in all the points
beyond $7$ kpc. 
If the inclination is taken as a free parameter in the fits, 
the MOND curve and the
data agrees well for an inclination of $24^{\circ}\pm 5^{\circ}$.
Swaters et al.~(2010) argue that an inclination of $25^{\circ}$
to $30^{\circ}$ is consistent with the H\,{\sc i} morphology
because  this galaxy has an optical bar with a faint
surrounding disk.  

NGC 4861 is a irregular dwarf galaxy with a luminosity
$L_{B}=7.9\times 10^{8}L_{\odot}$ and 
shows no evidence for spiral structure
(van Eymeren et al.~2009b). 
This galaxy was studied in H\,{\sc i} by Wilcots et al.~(1996),
Thuan et al.~(2004) and van Eymeren et al.~(2009a,b).
For $\Upsilon_{\star}^{B}=0.3M_{\odot}/L_{\odot}$,
the total baryonic mass is $9.1\times 10^{8}M_{\odot}$.
It is one of the most inclined galaxies in our selected sample.
The optical inclination from HYPERLEDA is $90^{\circ}$ (Paturel et 
al~2003).
Thuan et al.~(2004) derived an inclination of $82^{\circ}$ for the
outermost H\,{\sc i} tilted ring,
whereas van Eymeren et al.~(2009a,b) estimated a mean H\,{\sc i} inclination
of $65^{\circ}\pm 5^{\circ}$ within $7$ kpc.
Using the latter inclination,
we infer $\Gamma_{\rm obs}=4.7$ at $7$ kpc.

Tilted-ring fits for ESO 245-G05 give an inclination of 
$54^{\circ}\pm 10^{\circ}$ (C\^ot\'e et al.~2000).
Its H\,{\sc i} mass is $2\times 10^{8} M_{\odot}$.  
Adopting $\Upsilon_{\star}^{B}=1$, the total baryonic mass is 
$4.4\times 10^{8}M_{\odot}$.
At the last measured radius of $3.5$ kpc, 
we find $\Gamma_{\rm obs}=3.0\pm 0.6$, 
which is a factor of $2$ smaller
than the predicted value by MOND. Whereas this is a potential galaxy
to test MOND, it is still premature to make any conclusion
given the large errors quoted on the inclination and its
uncertain kinematics due to the presence of a strong bar.

HoII is an irregular galaxy in the M81 group. 
Careful analyses of H\,{\sc i} observations of HoII have been 
carried out by three independent groups: Puche et al.~(1992),
Bureau \& Carignan (2002)
and Oh et al.~(2011). The H\,{\sc i} rotation curves derived by all these
authors are very similar.  Oh et al.~(2011) derived 
the stellar mass-to-light ratio in K-band, $\Upsilon_{\star}^{K}$,
for HoII.
With such a $\Upsilon_{\star}^{K}$-value (which we will refer to as the
nominal value), the stellar mass in the disk of HoII
is of $1.5\times 10^{8}M_{\odot}$. 
Given that the total mass in gas is  $9\times 10^{8}M_{\odot}$ 
(Bureau \& Carignan 2002),
the total (gas plus stars) baryonic mass in HoII is about
$10.5\times 10^{8}M_{\odot}$. 
With these empirical values, we infer $\Gamma_{\rm obs}=2\pm 0.4$ at a radial
distance of $7$ kpc. This 
is much smaller than the value predicted by MOND, which is about $19$. 

ESO 364-G029 
has a luminosity of $L_{B}=6.6\times 10^{8}L_{\odot}$ for a distance
of $D=10.8$ Mpc (Kouwenhoven et al.~2007). 
The models of Bell \& de Jong (2001) predict
a stellar mass-to-light ratio of $\sim 0.2$--$0.6$ in the blue band, implying
a stellar mass of $(1.5-4)\times 10^{8}M_{\odot}$.
The H\,{\sc i} distribution is mildly asymmetric and roughly follows 
the stellar brightness distribution. 
The H\,{\sc i} mass is $6.4\times 10^{8}M_{\odot}$ and the rotation
velocity, assuming $i=70^{\circ}$, reaches a value of $42$ km s$^{-1}$ 
at a distance of $7$ kpc.
Despite the asymmetric appeareance of the H\,{\sc i},
the rotation curve is fairly symmetric. 
The corresponding $\Gamma_{\rm obs}$-value  at $R=7$ kpc 
is $2\pm^{1.0}_{0.5}$.

We conclude that, at present, the galaxies NGC 4861, HoII and ESO 364-G029 
are promising targets to test MOND. 
For the remainder four galaxies,
further decreasing systematic uncertainties in these galaxies
could provide a strong test to the MOND framework.
In the next section, we will consider
the whole available rotation curve of the galaxies NGC 4861, HoII
and ESO 364-G029 to quantify
the difference between predicted circular velocities and measured
velocities. The galaxies HoII and ESO 364-G029 are expected to
be the most problematic because they present the lowest
$\Gamma$-values ($\Gamma_{\rm obs}\simeq 2$).  
 
\section{Rotation curves: NGC 4861, HoII and ESO 364-G029}
\label{sec:testcases}
\subsection{The galaxy NGC 4861}
\begin{figure*}
\includegraphics[width=170mm, height=80mm]{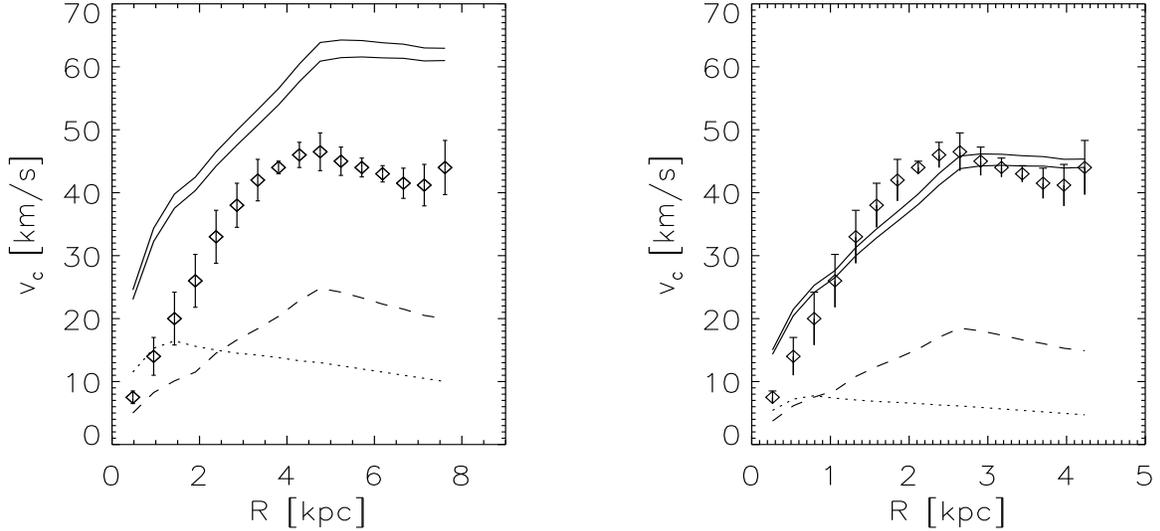}
 \caption{Rotation curve of NGC 4861 taken from van Eymeren et al.~(2009b),
together with the contributions of the stellar disk
(dotted lines) and gas (dashed lines) for $D=7.5$ Mpc and
$\Upsilon^{R}_{\star}=0.3$ (left panel) and for $D=4.2$ Mpc and
$\Upsilon^{R}_{\star}=0.12$ (right panel).
The full lines represent the rotation curve according to MOND prescription
using the simple $\mu$-function (upper solid curves) or the standard 
$\mu$-function (lower solid curves). 
Here we take $a_{0}=1.2\times 10^{-8}$ cm s$^{-2}$.
}
 \label{fig:model_NGC4861}
 \end{figure*}

The left panel of Figure \ref{fig:model_NGC4861} shows the observed 
rotation curve of NGC 4861
from van Eymeren et al.~(2009b) together with the predicted MOND
rotation curve (solid lines), for the simple and the standard
interpolating functions. 
The adopted distance was $D=7.5$ Mpc.
We have modelled only the inner $7.5$ kpc from
the dynamic center because beyond this radius the observed
rotation curve is affected by uncertainties caused by the sparsely
filled tilted rings. For the stellar disk, we have assumed 
$\Upsilon^{R}_{\star}=0.3M_{\odot}/L_{\odot}$. We see that 
MOND overestimates the observed rotation velocities at any radius.

If the distance and $\Upsilon_{\star}^{R}$ are left
as free parameters, 
an acceptable fit is obtained for 
$D=4.2$ Mpc and $\Upsilon_{\star}^{R}=0.12$ 
(see Fig.~\ref{fig:model_NGC4861}). 
The distance to this galaxy has been determined
with different methods and all gives $D>7$ Mpc.  
Using the Virgocentric infall model of
Schechter (1980) with parameters $\gamma=2$, $v_{\rm Virgo}=976$ km s$^{-1}$,
$w_{\odot}=220$ km s$^{-1}$ and a Virgo distance of $15.9$ Mpc,  
a distance of $10.7$ Mpc is derived for NGC 4861 
with an uncertainty of $\sim 20\%$ (Heckman et al.~1998;
Thuan et al.~2004).
Thus, it is unlikely that changes in distance
alone can explain the discrepancy between the predicted and the observed
rotation curves.

The MOND fit to the rotation curve can be improved if the inclination
of the galaxy and $\Upsilon^{R}_{\star}$ are left free in the fit. 
If, instead of
adopting the nominal inclination $i=65^{\circ}$ derived by van Eymeren 
et al.~(2009b), we use an inclination of $i=43^{\circ}$,  
the MOND curve and the derived rotation curve are consistent. 
However, the change required in inclination is much larger than the associated
uncertainty $\sim 5^{\circ}$, indicating that this inclination is 
unlikely.

\subsection{The galaxies HoII and ESO 364-G029}
\subsubsection{A comparative study}

\begin{sidewaystable}[h]\caption{Comparison of the relevant parameters of the selected galaxies}
\centering
\begin{tabular}{c c c c c c c c c c c c}\hline
{Name} & {$D$} & {$M_{B}$} & $L_{B}$ & {$\left<i\right>$} & {$M_{\rm gas}$} & 
$\Upsilon_{\star}^{B}$ & {$M_{\rm
bar}$} & $V_{\rm rot}$ & $\Gamma_{\rm obs}$ & $\Gamma_{M}$ & Source    \\[1ex]
{}&Mpc &  & $10^{8}L_{\odot}$ & deg & $10^{8}M_{\odot}$ && $10^{8}M_{\odot}$ &km s$^{-1}$    \\ [1ex]
\hline
E215?G009 & 4.2 & -12.9 & 0.23 & $36^{\circ}$ & $7.1$ & 0.5-2 
& 7.2--7.6 & $51\pm 5$ 
at $7$ kpc & $5.6\pm^{1.4}_{1.2}$  & $10.7\pm 2$ & 1 \\ [1ex]
   & 5.25   & -13.4 & 0.36  & $28^{\circ}$ & $11.1$ & 0.5--2 &  11--12 & $75\pm^{18}_{15}$ at $8$ kpc & $9.0\pm^{5.0}_{3.5}$& $6.1\pm^{3.0}_{1.8}$ & 2\\[1ex]
NGC 3077 & 3.8 & -17.75 & 19.6 & $46^{\circ}$ & 12.3 & 0.2--0.6 & 16--24 & $62\pm^{20}_{15}$ at $6$ kpc & $3.0\pm^{3.7}_{1.5}$ & $6.5\pm^{4.0}_{2.4}$ & 3,4\\
NGC 2366 & 3.4 & -17.17 & 11.5 & $63^{\circ}$ & 9.1 & 0.15--0.3 & 10.8--12.5 & $60\pm 16$ at $7$ kpc & $5.4\pm^{3.7}_{2.7}$ & $8.0\pm^{6.0}_{2.6}$ & 
2,3,5 \\
UGC 4173 & 18.0 & -16.50 & 6.2 & $40^{\circ}$ & 34.0 & 0.5--1 & 37--40 & $48\pm 5$ at $9$ kpc & $2.8\pm 0.7$ & $15\pm^{3.5}_{2.5}$ & 6 \\
NGC 4861 & 7.5 & -16.76 & 7.9 & $65^{\circ}$ & 6.7 & 0.15--0.3 & 8.0--9.1 & 
$46\pm^{5}_{2}$ at 7 kpc 
& $4.7\pm^{1.3}_{0.7}$ & $12.9\pm^{1.0}_{2.0}$ & 2,7 \\
E245-G05 & 2.5 & -15.0 & 1.6 & $54^{\circ}$ & $2.8$ & 0.7--1.5 & 3.9--5.2 & $43\pm 1$ at $3.5$ kpc 
& $3.0\pm 0.6$ & $7.8\pm 0.3$ & 8 \\
E364-G029 & 10.8 & -16.6 & 6.6 & 70$^{\circ}$ & $9.0$ & 0.2--0.6 &10--16 & 
$42\pm 3$ at 7 kpc& $2.0\pm^{1.0}_{0.5}$ & $15.0\pm ^{2.5}_{1.5}$ & 9\\
HoII & 3.4 & -16.87 & 8.7 & $49^{\circ}$ & 9.0  & 0.17 & 10.5 & $37\pm 4$ at 7 kpc & $2.0\pm 0.4$ & $19.5\pm^{4.5}_{3.5}$ & 3,5\\

\hline
\end{tabular}
\medskip\\
Column (1): Name of the galaxy. Column (2): Adopted distance. Column (3):
Absolute B magnitude. Column (4): Total blue-band luminosity.
Column (5): Average value of the inclination.
Column (6): Total mass in gas. Column (7): Stellar mass-to-light
ratio in the blue band. 
Column (8): Total baryonic mass. Column (9): Rotation velocity.
Column (10) and (11): $\Gamma_{\rm obs}$ and $\Gamma_{M}=1/\mu(g/a_{0})$.
Column (12): References, 1: Warren et al.~2004; 2: van Eymeren et al.~2009a;  
3: Walter et al.~2008; 4: Martin 1998; 5: Oh et al.~2011; 
6: Swaters et al.~2010; 7: van Eymeren et al.~2009b; 
8: C\^ot\'e et al.~2000; 9: Kouwenhoven et al.~2007.
\label{table:parameters}
\end{sidewaystable}

The galaxies HoII and ESO 346-G029 have similar baryonic mass but
the amplitudes beyond galactocentric distances of  $R=5$ kpc are
rather similar.
In the case of HoII, the method of Oh et al.~(2011) minimizes the effect of
non-circular motions. The rotation curve of HoII was corrected for 
asymmetric drift whereas this correction was not made for 
ESO 364-G029. 
Asymmetric drift corrections would cause a boost 
of a few km s$^{-1}$ in the circular velocities of these galaxies.

Remind that the asymptotic velocity is 
defined as $(GM_{\rm bar}a_{0})^{1/4}$, with
$M_{\rm bar}$ the baryonic mass. 
Using the estimates of the baryonic masses given in 
Table \ref{table:parameters} 
and $a_{0}=1.2\times 10^{-8}$ cm s$^{-2}$,
MOND predicts an asymptotic velocity of $63$ km s$^{-1}$ for HoII
and $66$ km s$^{-1}$ for ESO 364-G029.
The predicted rotational speed is in excess by $25$ km s$^{-1}$
in HoII and ESO 364-G029. In the following, 
we will concentrate on the case of HoII because
the rotation curve has a much better spatial resolution than
in ESO 364-G029. In addition, the irregular and lopsided morphology
of ESO 364-G029 could induce systematic biases if one assumes
axisymmetry.

\subsubsection{MOND in HoII}
\label{sec:MONDHoII}
Since HoII is embedded in the external gravitational field of M81 group, 
one has to quantify the external field effect (EFE) described in
Bekenstein \& Milgrom (1994), by comparing the internal and external
accelerations.
Assuming that the M81 group is bound, the external acceleration is
$0.7\times 10^{-10}$ cm s$^{-2}$ (Karachentsev et al.~2002), which
is much smaller than its 
internal accelerations ($\sim 6\times 10^{-10}$ cm s$^{-2}$ and 
$2\times 10^{-10}$ cm s$^{-2}$ at $R=7$ kpc and $R=20$ kpc, respectively).
Thus, EFE should be small at $R<10$ kpc.

\begin{figure*}
\includegraphics[width=160mm, height=65mm]{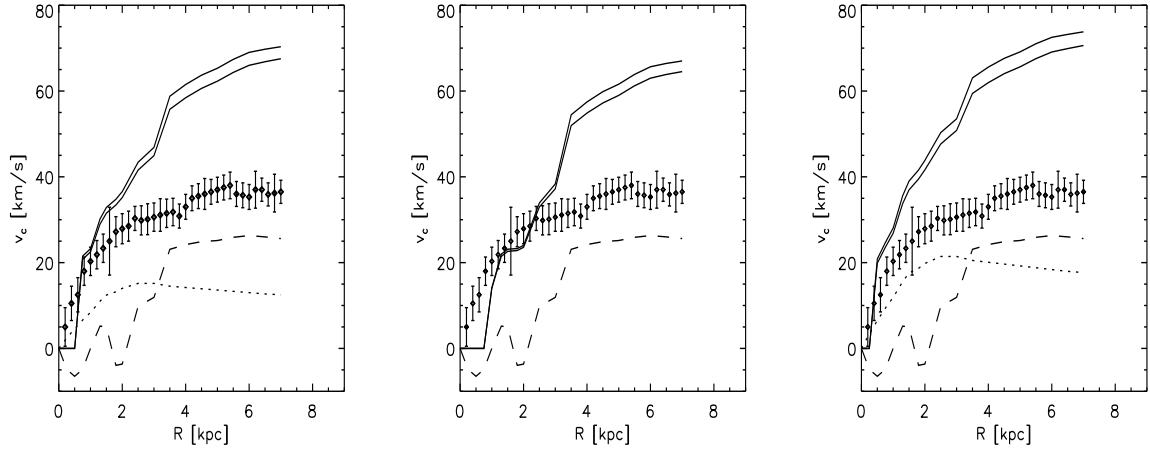}
 \caption{Rotation curve of HoII from Oh et al.~(2011)
together with the contributions of the stellar disk
(dotted lines) and gas (dashed lines) for the nominal
$\Upsilon_{\star}$-value (left panel), minimum disk plus gas
(central panel) and twice the nominal $\Upsilon_{\star}$-value.
The full lines represent the rotation curve according to MOND prescription
using the simple $\mu$-function (upper solid curves) or the standard 
$\mu$-function (lower solid curves). 
MOND is unable to provide the amplitude of the 
rotation curve. Here we take $D=3.4$ Mpc and
$a_{0}=1.2\times 10^{-8}$ cm s$^{-2}$.
}
 \label{fig:models_HoII}
 \end{figure*}

\begin{figure}
  \includegraphics[width=160mm,height=130mm]{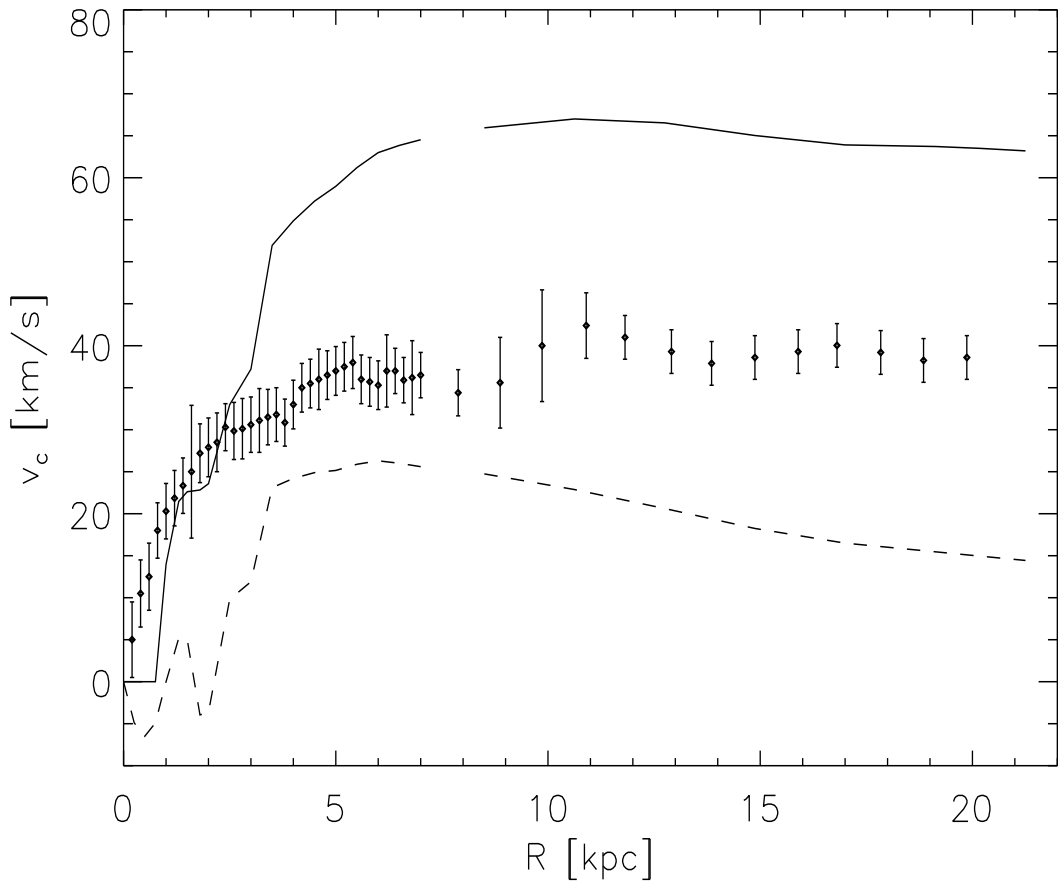}
\vskip 0.4cm
 \caption{Combined rotation curve of HoII as measured by Oh et al.~(2011) 
(at $R<7$ kpc) and by Bureau \& Carignan (2002) at $R>7$ kpc.  
The dashed line represents the contribution to the
rotation curve of the gas disk.  The solid line represents the 
MONDian rotation curve in the ``minimum disk$+$gas'' assumption,
using the standard $\mu$-function and $a_{0}=1.2\times 10^{-8}$ cm s$^{-2}$.
All the variables have been rescaled for the adopted distance of
$D=3.4$ Mpc.
}
 \label{fig:hybrid}
 \end{figure}

Figure \ref{fig:models_HoII} shows the predicted HoII rotation curve 
in MOND under various
$\Upsilon_{\star}^{K}$ assumptions [the nominal value as derived
by Oh et al.~(2011), ``minimum disk plus gas'', and twice the nominal
value]. The ``minimum disk plus gas'' includes the gas component and uses
the minimum value of $\Upsilon_{\star}$ compatible with the requirement 
that the theoretical circular velocity must be positive and 
reasonably smooth.
The discrepancy between the observed and the predicted rotation
curves is apparent.  
The effect of varying $\Upsilon_{\star}$ or the
interpolating function on the MOND circular velocity at the outer disk
is small, less than $10$ km s$^{-1}$ at $R=7$ kpc. 
For illustration, Figure \ref{fig:hybrid} shows the combined 
rotation curve from Bureau \& Carignan (2002)
and Oh et al.~(2011) for the updated HoII distance of $3.4$ Mpc. 
Note that at $R>10$ kpc, velocities were only measured in the approaching side.

The value of $a_{0}$ is universal but needs to be fixed from
observations.
Even adopting a value of $a_{0}$ at the lower end of the
best-fit interval derived by Begeman et al.~(1991)
and Gentile et al.~(2011), $a_{0}=0.9\times 10^{-8}$ cm s$^{-2}$,
the MOND circular rotation speed is only a few km s$^{-1}$ slower. 

\begin{figure}
  \includegraphics[width=160mm,height=130mm]{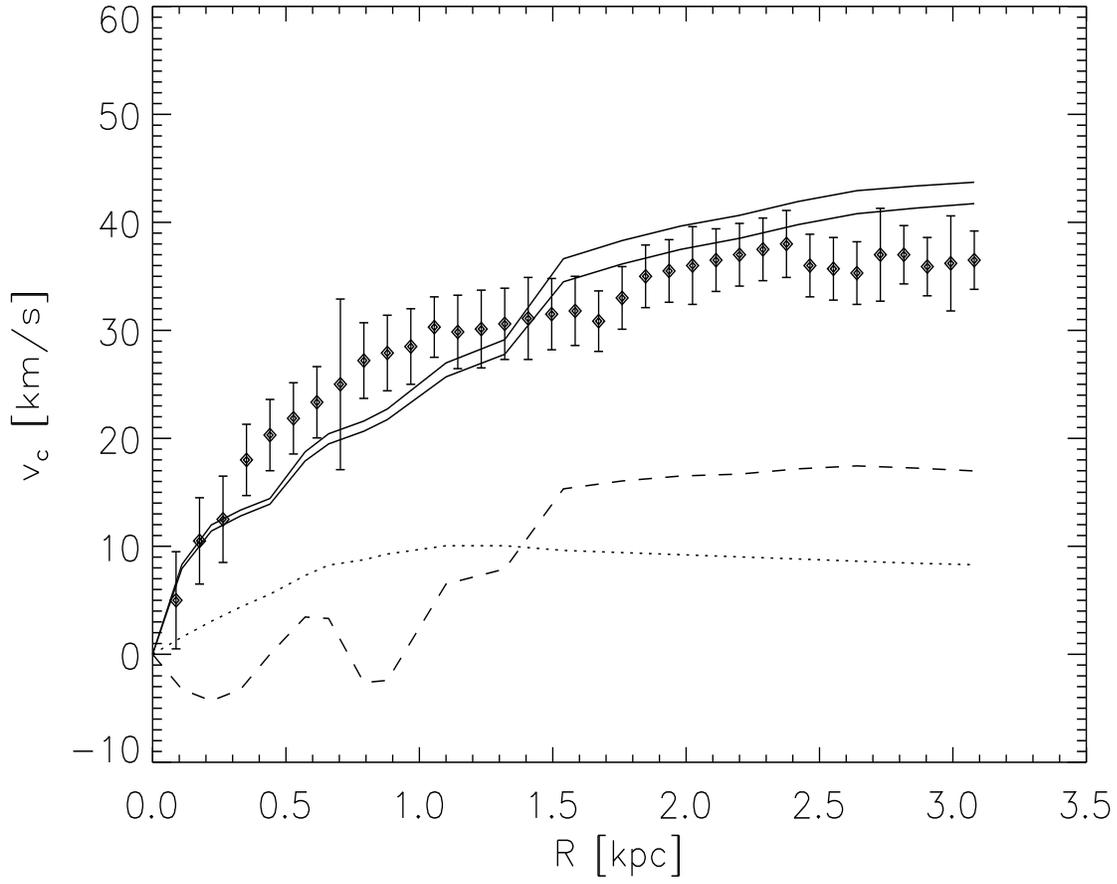}
 \caption{HoII rotation curve in MOND 
adopting the standard $\mu$-function with $a_{0}=0.9\times
10^{-8}$ cm s$^{-2}$ and a distance to the galaxy of $1.5$ Mpc, a factor $2.3$
closer than the nominal distance.
The key to lines is the same as in Fig.~\ref{fig:models_HoII}.  
}
 \label{fig:closer}
 \end{figure}

In the MOND prescription, 
the amplitude of the rotation curve can be accounted for 
by adopting a distance to HoII of $1.5$ Mpc and
$a_{0}=0.9\times 10^{-8}$ cm s$^{-2}$ (see Figure \ref{fig:closer}).
Given that the uncertainty in the distance is of $0.4$ Mpc (Karachentsev
et al.~2002), this likely
indicates that MOND cannot be made compatible with that rotation curve
by a reasonable adjustment of galaxy's distance.

A more delicate issue is the error resulting from the uncertainty in 
the inclination of the galaxy. 
The optical inclination from LEDA is of $45^{\circ}$ (Paturel et al.~2003).
Here, we have used a global inclination
of the  H\,{\sc i} disk of $40^{\circ}$ in the inner parts (Oh et al.~2011), 
and $84^{\circ}$ for tilted rings at $R>12$ kpc (Bureau \& Carignan 2002).
It turns out that if the inclination is taken as a free parameter,
a mean inclination of $25^{\circ}$ would yield a circular velocity
of $\sim 60$ km s$^{-1}$ at $R=7$ kpc. 
In a recent posting during the course of submitting this paper
and motivated by our preprint arXiv:1105.2612,
Gentile et al.~(2012) re-analyze the rotation curve in HoII
by modelling its H\,{\sc i} data cube and find that the inclination
is much closer to face-on than previously derived.
At this lower inclination, the rotation velocity becomes commensurate
with what is expected from MOND. 

In the very outer disk, at $R>12$ kpc,
an inclination of $45^{\circ}$ is required to reconcile
MOND with observations. This value is far lower than
the one derived by Bureau \& Carignan (2002) fitting tilted ring
models ($84^{\circ}$).
We conclude that more accurate determinations of HoII inclination
in the outer parts will provide a more definitive test to MOND.

\section{Final remarks and conclusions}
\label{sec:conclusions}
MOND predicts a tight correlation between
the asymptotic circular velocity and the total baryonic mass of the galaxy. 
Gas-rich dwarf galaxies are an interesting and unique test of modified
theories of gravity. Their internal accelerations are below $a_{0}$
and the stellar mass in these galaxies is not important in the budget 
of total mass. 

For a large sample of gas-rich dwarf galaxies with
rotation velocities between $35$ and $80$ km s$^{-1}$, we have computed 
the parameter $\Gamma_{\rm obs}$, 
defined as the ratio of real to Newtonian accelerations, and the
predicted value in MOND $\Gamma_{M}$.
We found that five galaxies (UGC 4173, HoII, ESO 245-G05,
NGC 4861 and ESO 364-G029) have $\Gamma_{\rm obs}<\Gamma_{M}$. 
For these galaxies, MOND overestimates the presently available rotation 
speeds, and hence, they cannot be fitted by arbitrarily increasing the 
mass-to-light ratio of the stellar component or by assuming additional
undetected matter. 

An amplitude of the rotation curve lower than
expected could be caused by an overestimate of either the inclination or the
distance to the galaxy.  
In order to quantify these effects, we have focused on two galaxies:
NGC 4861 and HoII. 
For these galaxies, we find that the discrepancy between
the observed and the predicted rotation curves in MOND
cannot be a consequence of adopting an incorrect distance
because it is unlikely that the distances are uncertain by that much.

An inclination of $25^{\circ}$, instead of $40^{\circ}$, is 
required to make the rotation curve of HoII compatible with MOND.
It is clear that the main source of systematic uncertainties in
the determination of the amplitude of the rotation curve in HoII
is its low inclination. 
The strategy is to look for a galaxy
as similar as possible to HoII but with a higher inclination to
reduce geometrical uncertainties.
We found that NGC 4861 and ESO 364-G029 
rotate at a similar velocity than HoII, have similar baryonic
mass but their inclinations are significantly larger. 
In the case of NGC 4861, the change required in inclination to fit
the observed rotation curve in MOND is much larger than the associated
uncertainty.
The galaxy ESO 364-G029 is very interesting because MOND
severely overpredicts by $50\%$ the observed circular velocity
at the outer parts of the disk.
Contrary to HoII, it has a comfortable inclination of $\sim 70^{\circ}$.
We conclude that deeper photometric and H\,{\sc i} 
observations of ESO 364-G029, together with further decreasing systematic
uncertainties, may provide a strong test to MOND.

\acknowledgements
We would like to thank Elias Brinks for his encouragement to make
these results public, Margarita Rosado for helpful discussions
and the anonymous referee for a useful report. 
The authors made use of THINGS
`The H\,{\sc i} Nearby Galaxy Survey' (Walter et al.~2008).
This research hasmade use of
the NASA/IPAC Extragalactic Database (NED) which is operated
by the Jet Propulsion Laboratory, California Institute of Technology,
under contract with the National Aeronautics and Space Administration.
This work was supported by the following projects: CONACyT 
165584, PAPIIT IN106212 and SIP-20121135.

{}

\end{document}